\documentclass[prb,twocolumn,superscriptaddress,nobibnotes]{revtex4}

\usepackage{graphicx}
\usepackage{epstopdf}
\usepackage{dcolumn}
\usepackage{bm}
\usepackage{times,mathptmx}
\usepackage{color}
\usepackage{multirow}
\usepackage{enumitem}
\usepackage{chngpage}

\begin{document}

\title{\bf Cavity optomechanical transduction sensing of single molecules}
\author{Wenyan Yu}
\thanks{These authors contributed equally to this work.}
\affiliation{Department of Electrical and Computer Engineering, University of Victoria, Victoria, BC V8P 5C2, Canada}
\author{Wei C. Jiang}
\thanks{These authors contributed equally to this work.}
\affiliation{Institute of Optics, University of Rochester, Rochester, NY 14627, USA}
\author{Qiang Lin}
\email{qiang.lin@rochester.edu}
\affiliation{Institute of Optics, University of Rochester, Rochester, NY 14627, USA}
\affiliation{Department of Electrical and Computer Engineering, University of Rochester, Rochester, NY 14627, USA}
\author{Tao Lu}
\email{taolu@ece.uvic.ca}
\affiliation{Department of Electrical and Computer Engineering, University of Victoria, Victoria, BC V8P 5C2, Canada}

\begin{abstract}
We report narrow linewidth optomechanical oscillation of a silica microsphere immersed in a buffer solution. Through a novel optomechanical transduction sensing approach,  single 10-nm-radius silica beads and Bovine serum albumin (BSA) protein molecules with a molecular weight of $66$ kDalton were detected. This approach predicts the detection of $3.9$ kDalton single molecules at a signal-to-noise ration above unity.
\end{abstract}

\maketitle
\section{Introduction}
Sensitive detection of a single nanoparticle/molecule is essential for many applications ranging from medical diagnostics, drug discovery, security screening, to environmental science. In the past decades, a variety of approaches have been developed, among which optical detection based on high-Q microcavities has shown significant advantages for its high sensitivity and label-free operation~\cite{Serpenguzel:95,Arnold08, Fan2008, Boyd:01,Lu12042011,Bowen_Back_scatter,ADOM:ADOM201400322,Lu_Vahala_Split_Systems,Zhu_Yang_On_chip_microresonator}.  Here, by taking advantage of the intriguing optically induced spring in a silica microsphere, we demonstrate that a quivering cavity can enhance the sensing resolution by orders of magnitude compared with conventional approaches, which allows us to detect single Bovine Serum Albumin (BSA) proteins with a molecular weight of $66$~kDalton at a signal-to-noise ratio (SNR) of $16.8$. This unique approach opens up a distinctive avenue that not only enables biomolecule sensing and recognition at individual level, but also exhibits great potential for studying/manipulating mechanical properties of individual biomolecules and their interactions.

\begin{figure}[hbtp]
  \includegraphics[width=0.45\textwidth]{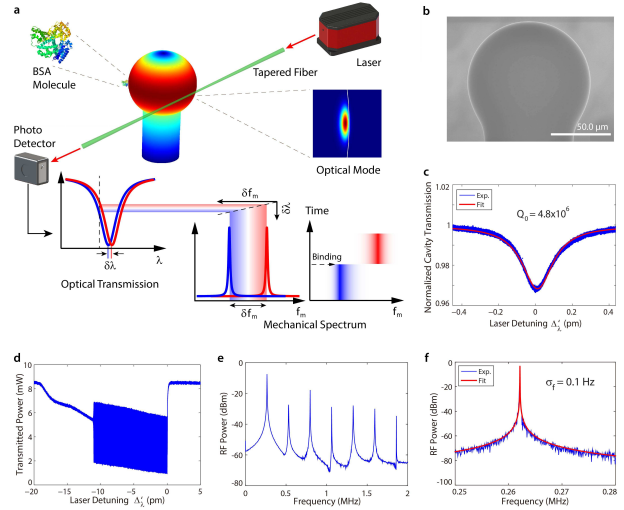}
   \caption{(a) Schematic illustrating the sensing mechanism. A protein molecule bound to an optomechanically oscillating microsphere yields an optical resonance shift $\delta\lambda$, which is transduced to a mechanical frequency shift $\delta{\rm f_m}$. The color map on the microsphere shows the radial breathing mechanical mode simulated by the finite element method. (b) A scanning electron microscopic (SEM) image of a fabricated silica microsphere. (c) The optical transmission spectrum of the microsphere immersed in DPBS, at a probe laser wavelength of 974~nm, with experimental data in blue and theoretical fitting in red. The input power is maintained low enough to characterize the intrinsic optical property of device, which exhibits an intrinsic optical Q of $4.8{\times}10^6$. (d) The optical transmission spectrum at an input laser power of $8.5$~mW. The coherent OMO was excited with a threshold power of 3.0~mW dropped into the cavity. (e) An example of the power spectral density of the cavity transmission. The fundamental oscillation frequency is located at 262~kHz, with 6 high-order harmonics clearly visible on the spectrum. (f) The detailed spectrum of the fundamental oscillation tone, with experimental data in blue and theoretical fitting in red. The OMO exhibits a full-width at half maximum of 0.1~Hz, corresponding to an effective mechanical Q of $2.6\times 10^6$.
   }\label{Fig_simulation}
\end{figure}

Binding of a particle to a high-Q optical microcavity perturbs the cavity mode at a resonance wavelength of $\lambda_0$, resulting in a cavity resonance shift of $\delta{\lambda}$ which in turn changes the cavity transmission. This mechanism underlies the majority of current microcavity sensors, with a sensing resolution dependent critically on the optical quality factor (Q)~\cite{Fan08}. To date, the highest resolution reported is a resonance shift of $(\delta\lambda/\lambda_0)= 3{\times}10^{-10}$ achieved with an optical Q of one hundred million at a visible wavelength in an aqueous environment~\cite{Lu12042011}, which, however, is still larger than that induced by a single protein binding event~\cite{Fan2008}. Consequently, detection of single protein molecule requires incorporating a plasmonic nanoantenna on the microcavity to enhance the resonance wavelength shift~\cite{Min2009,Shopova2011,Dantham_Label,Baaske_Vollmer_Single_molecule}, at a price of significant reduction of effective detection area.

On the other hand, the optical wave cycling inside the microcavity is able to produce a radiation pressure that interacts with the mechanical motion of the device (Fig.~\ref{Fig_simulation}a). Such optomechanical coupling flourishes in profound physics that has been intensively explored in recent years, particularly in the context of quantum control of mesoscopic mechanical motion~\cite{Kippenberg2008, Marquardt09, Favero_Optomechanics, Thourhout_Optomechanical_NP,Aspelmeyer13}. When the laser wavelength is blue detuned to the cavity resonance, the optical wave can efficiently boost the mechanical motion above the threshold of regenerative oscillation~\cite{Kippenberg2008, Aspelmeyer13}, resulting in highly coherent optomechanical oscillation (OMO) with a narrow mechanical linewidth. Of particular interest is that the optical wave is able to produce an effective mechanical rigidity whose magnitude depends sensitively on the laser-cavity detuning~\cite{Aspelmeyer13}. Consequently, any tiny perturbation to the cavity resonance wavelength induced by particle/molecule binding would be readily transferred to the frequency shift of the mechanical motion, enabling an efficient transduction mechanism to amplify the resonance wavelength sensing. With the narrow linewidth of coherent OMO, the intriguing optical spring effect would thus offer an elegant approach for sensitive probing of cavity resonance variation, with a sensing resolution given by
\begin{eqnarray}
\left(\frac{\delta\lambda}{\lambda_0} \right)_{\rm min}= \frac{1}{\eta_{om} Q_m Q_t}, \label{DeltaLambda_res}
\end{eqnarray}
where $Q_m$ is the effective mechanical Q factor of OMO, $Q_t$ is the loaded optical Q, and $\eta_{om}$ represents the optomechanical transduction factor for sensing whose magnitude depends on laser-cavity detuning, with a value in the order of $\eta_{om}\sim{1}$. Equation~(\ref{DeltaLambda_res}) shows clearly that the sensing resolution scales not only with the optical Q of the cavity as what does in conventional microcavity sensors, but also with the effective mechanical Q of OMO. Consequently, in principle, the proposed cavity optomechanical transduction sensing is able to enhance the sensing resolution by about a factor of $Q_m$ compared with conventional approaches. As we will show below, the mechanical Q of coherent OMO in our device can reach a value above $10^6$, resulting in a sensing resolution enhanced by orders of magnitude that is well sufficient for single molecule detection.

As the optomechanical effect is intrinsic to a high-Q microcavity, the proposed approach does not rely on any specific external sensing element attached to the device ($e.g.$, a plasmonic nanoantenna~\cite{Shopova2011,Dantham_Label,Baaske_Vollmer_Single_molecule}), thus capable of fully utilizing the entire effective sensing area offered by a whispering-gallery microcavity which is more than five orders of magnitude larger than that of plasmonic devices. For the same reason, it does not rely on any gain medium, thus universal to different material platforms as long as the device has reasonably high optical Q. On the other hand, this approach is distinctive from the conventional micro-/nano-mechanical sensing~\cite{Craighead07, Roukes11} where the particle detection is realized by monitoring the mechanical frequency shift directly induced by the mass change from particle attaching, which, however, exhibits a minimal detectable mass of $(\delta m)_{\rm min} = \frac{2 m_{\rm eff}}{Q_m}$ that relies critically on the motional mass $m_{\rm eff}$ and the mechanical Q of the sensor. Although this mechanism was recently applied in combination with optical actuation and readout, it can only detect $2$-${\rm \mu{m}}$-diameter silica beads with a picogram resolution~\cite{6297435,Fan2014,PhysRevLett.108.120801}.
\begin{figure}[hbtp]
      \includegraphics[width=0.45\textwidth]{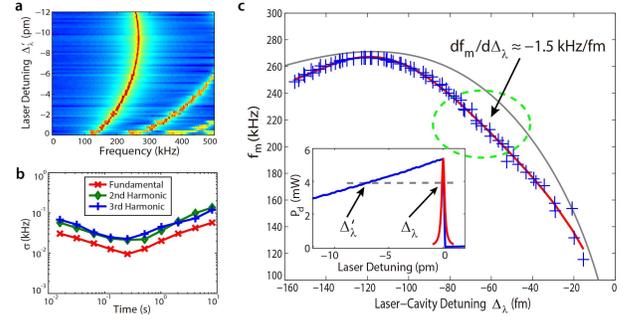}
  \caption{(a) Spectrogram of cavity transmitted signal as a function of laser wavelength detuning $\Delta^\prime_{\lambda}$ (see Fig.~\ref{Fig_simulation}d for the meaning of $\Delta^\prime_{\lambda}$), showing the detuning dependent mechanical frequency. The proportional frequency variations at the second and third harmonics are clearly visible. Every spectrum was averaged over $5$ traces acquired continuously. (b) The two-sample Allan deviations of the fundamental, second and third harmonic tones measured in DPBS in the absence of sensing particle, showing a minimum deviation of $9.5$~Hz at the fundamental oscillation tone. (c) The OMO frequency as a function of laser-cavity wavelength detuning. The blue crosses show the experimental data and the grey curve shows the theory. The red curve is a polynomial fitting to the experimental data. The dashed circle indicates the operating regime for the particle and molecule sensing, with a frequency tuning slope of $d{\rm f_m}/d\Delta_\lambda = -1.5~{\rm kHz/fm}$ at a laser-cavity detuning of $\Delta_\lambda = -70~{\rm fm}$. Inset: Recorded dropped optical power as a function of laser wavelength detuning. This curve was used to obtain the real laser-cavity wavelength detuning $\Delta_\lambda = \lambda_l - \lambda_0$ where $\lambda_l$ is the laser wavelength. }\label{Fig_detune}
    \end{figure}
\section{Results and discussions}
To verify the above sensing principle, we carried out experiments in a silica microsphere with a diameter of about 100~${\rm \mu m}$. The device exhibits an intrinsic optical Q as high as $4.8{\times}10^6$ at a wavelength around $974$~nm in the aqueous environment (Fig.~\ref{Fig_simulation}c), which is close to the theoretical limit. With such a high optical Q, the optical wave inside the microsphere produces a strong radiation pressure that efficiently actuates the radial breathing mechanical motion of the microsphere (Fig.~\ref{Fig_simulation}a). Consequently, by dropping an optical power of $3.0$~mW into the cavity (Fig.~\ref{Fig_simulation}d), we are able to boost the mechanical mode above the threshold even in the aqueous environment, resulting in coherent OMO at a frequency of $262$~kHz with a mechanical linewidth as narrow as $0.1$~Hz (Fig.~\ref{Fig_simulation}f), corresponding to an effective mechanical Q of $2.6\times 10^6$. As shown in Fig.~\ref{Fig_simulation}e, the significant optomechanical oscillation leads to a harmonic comb on the power spectrum of the cavity transmission, a feature of coherent OMO resulting from the nonlinear transduction of the optical cavity~\cite{Hossein-Zadeh2006, Xiong_Tang_Integrated_Resonators}.

In particular, the strong optical spring effect from the optomechanical coupling results in an OMO frequency sensitively dependent on the laser-cavity detuning (Fig.~\ref{Fig_detune}a). As shown in Fig.~\ref{Fig_detune}c, when the laser-cavity wavelength detuning decreases from $-150$~fm, the OMO frequency increases from $247$~kHz to a peak value of $267$~kHz, and then decreases quickly to about $115$~kHz when the laser wavelength is tuned close to the center of cavity resonance, with a tuning slope of $d{\rm f_m}/d\Delta_\lambda\approx-1.5$~kHz/fm at a laser-cavity detuning of $\Delta_\lambda \approx -70~{\rm fm}$. The observed optical spring follows closely the theoretical expectation (grey curve in Fig.~\ref{Fig_detune}c).

Such an optical spring corresponds to a sensitive \emph{optical-to-mechanical frequency transduction}, inferring that every $1$-fm cavity resonance wavelength shift induced by a particle binding event can be transduced to an OMO frequency change of about $1.5$~kHz that is about four orders of magnitude larger than the linewidth of optomechanical oscillation (Fig.~\ref{Fig_simulation}f). A detailed characterization of the Allan deviation of the OMO frequency (Fig.~\ref{Fig_detune}b) shows a minimum two sample deviation of $9.5$~Hz at the fundamental OMO frequency, implying a detection resolution of $\delta\lambda/\lambda_0\approx{6}\times{10}^{-12}$ in the device. This record resolution clearly shows the powerfulness of the demonstrated approach, which is more than $10^4$ times higher than a conventional microcavity sensor with the same optical Q~\cite{Fan08}.
It is even about $50$ times higher than that achieved with an optical Q of $10^8$ at a visible wavelength~\cite{Lu12042011}. Note that the harmonics of the optomechanical oscillation vary proportionally with the OMO fundamental frequency (Fig.~\ref{Fig_detune}a) and thus can also be applied for particle sensing. Although this does not improve the sensing resolution due to the same SNR of detection, in practice, the larger frequency shifts on the higher-order harmonics significantly facilitate the mechanical spectrum analysis (by allowing to use a coarser resolution bandwidth), which reduces considerably the excessive detection noises from consecutive multiple particles skimming by the cavity surface.
 \begin{figure}[hbtp]
      \includegraphics[width=0.45\textwidth]{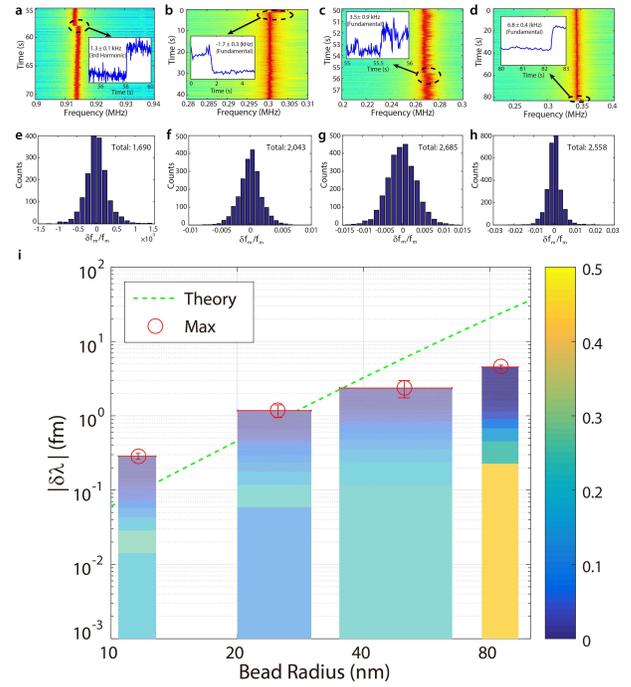}
  \caption{ (a)-(d) Typical mechanical spectrograms for the binding events of silica beads with average radii of $11.6$, $25$, $50$, and $85$~nm, where (a) shows that of third harmonic and (b)-(d) show those of the fundamental oscillation frequency. The insets show the detailed frequency steps. (e)-(h) The histograms of the normalized frequency steps $\delta {\rm f_m}/{\rm f_m}$. (i) The corresponding cavity resonance shifts induced by the particle binding as a function of bead radius. The color bars show the probability density functions of the recorded cavity resonance wavelength shifts induced by particle binding, where the bar width indicates the standard deviation of the bead size (provided by the manufacturer) and the color map indicates the magnitude of probability density. The red circles indicate the recorded maximum wavelength shifts of the cavity resonance. The dashed curve shows the theortical prediction~\cite{biosensing_Arnold_Vollmer_Shift}. }\label{Fig_steps}
  \end{figure}

To characterize the real sensing performance, we performed the sensing experiments on silica nanobeads with different diameters. We set the laser-cavity detuning at the operational point indicated within the dashed circle of Fig.~\ref{Fig_detune}c and delivered silica nanobeads diluted in Dulbecco's Phosphate-Buffered Saline (DPBS) around the microsphere. The particle binding events were recorded by searching the sudden changes of the oscillation frequency in the recorded spectrograms. Typical examples are shown in Fig.~\ref{Fig_steps}a-d for the nanobeads with radii of $11.6$, $25$, $50$, and $85$-nm, respectively. Here in the case of $11.6$-nm beads, the frequency steps were recorded at the third harmonic of the oscillation frequency while all others were obtained at the fundamental. As shown in Fig.~\ref{Fig_steps}a, a clear step of $1.3{\pm}0.1$~kHz (corresponding to $0.43{\pm}0.03$~kHz step at the fundamental oscillation tone) was observed at the time of $58$ second with an SNR of $13$. An increase of the oscillation frequency implies a red shift of the cavity resonance wavelength, which corresponds to a binding of a $11.6$-nm silica bead on the surface of the microsphere. Figures~\ref{Fig_steps}b-d show frequency steps of $-1.7{\pm}0.3$, $3.5{\pm}0.9$, and $6.8{\pm}0.4$~kHz, respectively, which correspond to the binding (positive frequency steps) or unbinding (negative steps) events of $25$-nm, $50$-nm, and $85$-nm beads.

To obtain the statistical properties of the binding events, we recorded a total number of $500121$, $521389$, $1335415$, and $758728$ spectra in sensing $11.6$-nm, $25$-nm, $50$-nm, and $85$-nm beads, respectively, among which frequency steps of 1690, 2043, 2685, and 2558 were captured with the SNR exceeding unity. Figures \ref{Fig_steps}e-h show the histograms of the normalized frequency steps, $\delta {\rm f_m}/{\rm f_m}$, which indicate maximum OMO frequency shifts of $\delta {\rm f_m}/{\rm f_m} = (1.4{\pm}0.4){\times}10^{-3}$, $(-7.8{\pm}1.5){\times}10^{-3}$, $(1.3{\pm}0.3){\times}10^{-2}$, and $(-2.3{\pm}0.6){\times}10^{-2}$, respectively, for beads with radii of $11.6$, $25$, $50$, and $85$~nm. We converted the recorded OMO frequency steps into the corresponding cavity resonance wavelength shifts $\delta \lambda$, with the transduction rate of $d{\rm f_m}/d\Delta_\lambda = - 1.5~{\rm kHz/fm}$. The probability density function of their absolute values are plotted as color bars in Fig.~\ref{Fig_steps}i, with maximum wavelength shifts of $|\delta \lambda|/\lambda_0 = 2.6{\times}10^{-10}$, $1.2{\times}10^{-9}$, $2.4{\times}10^{-9}$, and $4.6{\times}10^{-9}$, respectively, for the four bead sizes (red circles). For comparison, we also numerically estimated the expected maximum wavelength change as a function of bead radius (green dashed line)~\cite{biosensing_Arnold_Vollmer_Shift}. The experiment results agree with the theoretical predictions when the bead size is small ($11.6$-nm and $25$-nm). At larger bead radii ($50$-nm and $85$-nm), the experiment values are smaller than the theoretical predictions, which is because the optical Q starts to degrade at a large beam size. As the optical spring depends on both the laser-cavity detuning and the optical Q, the impact from
cavity Q change counteracts that from the cavity resonance wavelength shift, leading to a smaller shift of OMO frequency
  \begin{figure}[hbtp]
      \includegraphics[width=0.45\textwidth]{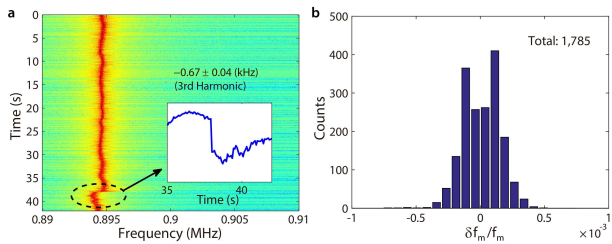}
  \caption{(a) A typical mechanical spectrogram recorded at the third harmonic of the oscillation tone, capturing the event of a BSA protein molecule detaching from the silica microsphere surface at $38$ second, with a clear frequency step (inset) of $-0.67{\pm}0.04$~kHz. (b) The histogram of the normalized frequency steps.}\label{Fig_BSA}
  \end{figure}

The ultrahigh detection sensitivity demonstrated on silica nanobeads readily implies the superior capability of sensing single protein molecules. To do so, we injected BSA diluted in DPBS around the microsphere sensor, with the concentration gradually increased form $1$~nM to $10$~nM. In the protein sensing experiment, the excessive noises from the unwanted molecules were significantly reduced, resulting in a detection noise level close to the DPBS background noise obtained from the Allan deviation measurement. The third order harmonic of the oscillation frequency was employed to monitor the frequency steps. As shown in Fig.~\ref{Fig_BSA}a, a maximum frequency step of $-0.67{\pm}0.04$~kHz was observed with an SNR of $16.8$, which corresponds to a $-0.22{\pm}0.01$~kHz at the fundamental oscillation tone. In total, we recorded 145407 spectra among which $1785$ frequency steps were captured. The histogram of the normalized frequency steps is plotted in Fig.~\ref{Fig_BSA}b, with the maximum step of $\delta {\rm f_m}/{\rm f_m}=(-7.6{\pm}0.4){\times}10^{-4}$ which corresponds to a cavity resonance shift of $|\delta \lambda|/\lambda_0 = 1.5{\times}10^{-10}$. This observation clearly proves the capability of sensing single BSA molecule with a molecule weight of 66~kDalton. By assuming the resonance shift is proportional to the mass (or equivalently, to the volume) of the protein~\cite{biosensing_Arnold_Vollmer_Shift}, we derive that our current setup is capable of detecting proteins as small as $3.9$~kDalton with an SNR above unity.

\section{conclusion}
The demonstrated single molecule detection now paves the foundation of ultra-sensitive cavity optomechanical transduction sensing. The sensing resolution can be further improved significantly in the future. For example, the minimal detectable OMO frequency shift in current devices is primarily limited by the laser frequency jitter in our experiment. With an OMO linewidth of only $\sim 0.1~{\rm Hz}$, we expect that the future adoption of a fine laser frequency locking circuitry can further improve the sensing resolution by $\sim 100$ times to around $\delta\lambda/\lambda_0\sim {10}^{-14}$. On the other hand, the optical Q can be increased to above $10^8$ if a visible laser is employed~\cite{Lu12042011}, which would further improve the sensing resolution by more than one order of magnitude. Moreover, a plasmonic structure can also be incorporated to enhance the cavity resonance shift. These future improvements would enable detecting small molecules and atoms with a mass down to sub-Dalton level, with a great potential for dramatically advancing the capability of biosensing to an unprecedented level.

In particular, as the molecule binding occurs during the coherent mechanical motion of the sensor, controlling the motion pattern of the coherent OMO (amplitude, phase, time waveform, etc.) may function as a unique paradigm to study/control the mechanical properties of molecule binding/unbinding. This, in combination with certain functionalization of the sensor surface~\cite{Arnold08} and with implementation of potentially versatile optomechanical motions~\cite{Aspelmeyer13}, may offer a unique multifunctional biomolecule toolbox that is not only able to observe cellular machineries at work, but also to selectively manipulate single-molecule interactions~\cite{Muller08}.

\noindent\textbf{Acknowledgements} The related contents are pending for patent application. W.Y. and T.L. acknowledge the support from NSERC Discovery and CFI LOF funds. W.C.J. and Q.L. acknowledge the support from the DARPA QuASAR program.

\noindent\textbf{Author Contributions} W.Y. fabricated the devices, prepared the samples, and performed the experiments. W.C.J. and Q.L. discovered the sensing principle and mechanism, developed the theory, and conducted the numerical simulation. T. L. conceived and designed the experiments, developed software to control the experiments, and processed the data. Q.L. and T.L. conceived the concept and supervised the project. W.Y., W.C.J., Q.L., and T.L. worked together on the result analysis/interpretation and manuscript preparation.

\noindent\textbf{Author Information} The authors declare no competing financial interests.  Correspondence and requests for materials should be addressed to T.L. ($taolu@ece.uvic.ca$) and Q.L. ($qiang.lin@rochester.edu$).

\end{document}